# Pulse Pileup Rejection Methods Using a Two-Component Gaussian Mixture Model for Fast Neutron Detection with Pulse Shape Discriminating Scintillator

Andrew Glenn*, Qi Cheng**,
Alan D. Kaplan**, Ron Wurtz*

* *Nuclear and Chemical Sciences Division, Lawrence Livermore National Laboratory,
Livermore, CA, United States.*
** *Computational Engineering Division, Lawrence Livermore National Laboratory,
Livermore, CA, United States*
*Corresponding author: wurtz1@llnl.gov, 925-423-8504,
Lawrence Livermore National Laboratory L-211, 7000 East Avenue, Livermore CA 94550*

**Abstract:** Pulse shape discriminating scintillator materials in many cases allow the user to identify two basic kinds of pulses arising from two kinds of particles: neutrons and gammas. An uncomplicated solution for building a classifier consists of a two-component mixture model learned from a collection of pulses from neutrons and gammas at a range of energies. Depending on the conditions of data gathered to be classified, multiple classes of events besides neutrons and gammas may occur, most notably pileup events. All these kinds of events are anomalous and, in cases where the class of the particle is in doubt, it is preferable to remove them from the analysis. This study compares the performance of several machine learning and analytical methods for using the scores from the two-component model to identify anomalous events and in particular to remove pileup events. A specific outcome of this study is to propose a novel anomaly score, denoted G, from an unsupervised two-component model that is conveniently distributed on the interval [-1,1].

## 1. Introduction

This work describes a method to reject anomalies using the scores based on likelihoods provided by an unsupervised two-component statistical model trained with data consisting almost entirely of two classes of digitized pulses. In this case, the statistical model is a Gaussian Mixture Model (GMM), the training data are low-count-rate (low pileup) Cf-252 pulses, and the components and scores correspond to neutrons and gammas. The anomalies of primary interest are pileup pulses, with the specific concern of pileup gammas that would be misclassified as neutrons. For this study, we treat the neutrons as the specific desirable class, and ignore distinctions within the leftover class.

## 2. Pulse Shapes

*2.1 Neutrons and gammas*

In conventional PSD materials, fast neutrons interact with protons which recoil in the material and ionize molecules densely along a short path, while gammas interact with electrons which recoil in the material and ionize molecules sparsely along a longer path. Owing to density-dependent de-excitation processes, the de-excitation times are different for the two kinds of particles, and the neutron scintillation pulses are broader in time than the gamma pulses [1,2,3,4].

As observed by electronics sensitive to approximately ten nanoseconds, a scintillation pulse for both kinds of particles has a fast and a slow component that each decay in time. The neutron pulse is broader than a gamma pulse because the proportion of the slowly-decaying component to the faster component is greater for a neutron. Careful pulse-by-pulse shape analysis can discriminate between these two kinds of pulses, and many pulse classification methods have been proposed and implemented, using analog [1] and digital [5] electronics.

*2.2 The shapes of pileups*

If two particles interact nearly simultaneously in the scintillating volume, then their scintillation pulses overlap in time. Such overlapping pulses, called "pileups", confuse the discrimination between neutrons and gammas. A second small pulse, delayed a little bit from a preceding gamma pulse, causes the sum of the two pulses to be apparently broadened and look more like a single pulse from a neutron. The result is that, in conditions of high event rate, the probability of a gamma pulse being broadened increases and thus the rate of gamma pulses misidentified as neutrons will increase.

## 3. PSD Methods

For fast neutron detection, the most basic pulse discrimination case consists of distinguishing a single pulse from the absence of a pulse, the more particular case consists of discriminating neutron pulses from gammas, and the most subtle case involves discriminating anomalous pulses from neutrons and gammas. For the first case, solutions used by the authors rely on trigger algorithms implemented in commercially-available digital pulse processing hardware and firmware [6]. The authors covered the second case in a recent series of papers [7,8,9,10,11]. This paper discusses the third case: trigger-collected pulses which exhibit shapes unlike those from either neutrons or gammas.

Any kind of pulse classification relies on a universal principle: one kind of pulse is known to exhibit one kind of shape, other kinds are known to exhibit other kinds of shapes. The shapes are quantified using data collected under known conditions, commonly called "training data". A digitized waveform is pre-processed and compared against these quantified shapes (possibly including variance), and a comparison parameter is obtained for the waveform for each kind of known shape. The waveform's comparison parameters are then combined into a score or set of scores, and those scores are used to discriminate between neutrons, gammas, and anomalies. The methods of pre-processing, comparison, and discrimination form a large menagerie of approaches proposed and implemented over the years [5].

*3.1 GMM for non-pileup gamma-neutron*

In previous papers [7,8,10], working under a practical constraint to use only unsupervised methods, the authors showed that a two-component GMM is a straightforward and effective method of relying on unlabeled training data consisting of neutrons and gammas to obtain a score to discriminate neutrons from gammas. GMM finds clusters in the multivariate space of observed features of the training data that contains both neutrons and gammas. If the neutrons and gammas are sufficiently well-separated in feature space, then neutrons and gammas are the two components found by the GMM. To classify a new event, the GMM provides two scores describing consistency with each of the components, and the two scores can be combined into a final single score to discriminate between neutrons and gammas.

The component scores can be expressed a number of different ways. Each Gaussian component $j$ is assigned a mean vector $\boldsymbol{\mu}_j$ and a covariance matrix $\boldsymbol{\Sigma}_j$. Then, for any measured pulse, variance-normalized distances — called Mahalanobis distances $D_j$ [12] — can be obtained between the pulse's feature vector $z$ and the means of each of the fitted Gaussians:

$$D_j(\boldsymbol{z}) = D(\boldsymbol{z}; \theta_j) = \sqrt{(\boldsymbol{z}-\boldsymbol{\mu}_j)' \boldsymbol{\Sigma}_j^{-1} (\boldsymbol{z}-\boldsymbol{\mu}_j)}$$

where $\theta_j = (\boldsymbol{\mu}_j, \boldsymbol{\Sigma}_j)$. The Mahalanobis distance appears in the likelihood function $f$ for each cluster $j$:

$$f_j(\boldsymbol{z}) = f(\boldsymbol{z}|\theta_j) = \frac{1}{\sqrt{(2\pi)^d |\boldsymbol{\Sigma}_j|}} \exp\left(-\tfrac{1}{2} D_j^2(\boldsymbol{z})\right) \quad (1)$$

where $d$ is the number of dimensions of the feature space. And the two likelihoods appear in the expression (likelihood ratio) of the likelihood that an observed event $z$ is consistent with a model of the neutron pulse shape,

$$\begin{aligned} P_n(\boldsymbol{z}) &= P(class = n|\boldsymbol{z}; \theta_n, \theta_\gamma, \alpha_n, \alpha_\gamma) \\ &= \frac{\alpha_n f_n(\boldsymbol{z})}{\alpha_n f_n(\boldsymbol{z}) + \alpha_\gamma f_\gamma(\boldsymbol{z})} \end{aligned} \quad (2)$$

where $n$ and $\gamma$ denote neutrons and gammas, and the alphas are mixing coefficients from the expected proportions of neutrons and gammas. It is important that the set of training data be collected, inspected, and cleaned such that contamination from anomalous pulses or other phenomena is very low, so that only the two clusters corresponding to neutrons and gammas are present. There are several options for processing of the pulse data and GMM structure that may affect performance. Carefully quantified performance metrics guide the user to choose among these options. The authors selected a specific method of pre-processing the feature vectors that yielded the best neutron-gamma discrimination results for the data-collection setup described in Section 4.1 when scored using labeled data and evaluated with a ROC curve [7]. The user is also free to choose the most appropriate covariance structure for the GMM, and the authors selected a shared covariance matrix (same covariance matrix across the two components), and used only the diagonal entries (variances) to compute the Mahalanobis distance [8], as follows:

$$D_j(\boldsymbol{z}) = \sqrt{\sum_i \frac{(z_i - \mu_{j_i})^2}{\sigma_{j_{ii}}}}$$

where $\sigma_{j_{ii}}$ are the diagonal elements of $\boldsymbol{\Sigma}_j$. Depending on the completeness of the statistical model, a few different measures of similarity to a given condition are available. As described above, the probability of a pulse belonging to either of the components can be computed from the two means and covariances and two mixing coefficients, while the likelihoods in Equation (1) are computed from only the means and covariances.

Even when there is insufficient knowledge to compute probability, the likelihoods can be used to discriminate between the two classes. The optimal binary test between gammas and neutrons, in the Neyman–Pearson sense [13], is a likelihood ratio test between neutron and gamma pulse distributions. The log-likelihood ratio (LLR) is the difference of the two log-likelihoods, with a single log-likelihood:

$$\ln(f_j(\boldsymbol{z})) = -\tfrac{1}{2} D_j^2(\boldsymbol{z}) + K_j$$

where

$$K_j = \ln\left(\frac{1}{\sqrt{(2\pi)^d |\Sigma_j|}}\right)$$

And the log-likelihood ratio is:

$$LLR(\mathbf{z}) = \ln\left(f_n(\mathbf{z})/f_\gamma(\mathbf{z})\right)$$
$$= -\tfrac{1}{2}D_n^2(\mathbf{z}) + K_n - (-\tfrac{1}{2}D_\gamma^2(\mathbf{z}) + K_\gamma)$$
$$= -\tfrac{1}{2}(D_n^2(\mathbf{z}) - D_\gamma^2(\mathbf{z})) + K_n - K_\gamma$$

And so, because we selected a shared covariance matrix $K_n = K_\gamma$ for the covariance structure:

$$LLR(\mathbf{z}) = \tfrac{1}{2}(D_\gamma^2(\mathbf{z}) - D_n^2(\mathbf{z}))$$

Note that it is necessary to choose the numerator and denominator of the ratio in the LLR. The authors prefer to make LLR for the neutrons positive. If the vector of pulse samples $\mathbf{z}$ were exactly the mean neutron pulse, $\mathbf{z} = \boldsymbol{\mu}_n$, then $D_n^2(\mathbf{z} = \boldsymbol{\mu}_n) = 0$, and since $\boldsymbol{\mu}_n \neq \boldsymbol{\mu}_\gamma$ then $D_\gamma^2(\mathbf{z} = \boldsymbol{\mu}_n) > 0$, making $LLR(\boldsymbol{\mu}_n) > 0$, so we define the LLR as log of neutron likelihood over gamma likelihood.

We define one last pair of scores:

$$S_\gamma(\mathbf{z}) = -\tfrac{1}{2}D_\gamma^2(\mathbf{z}), \qquad S_n(\mathbf{z}) = -\tfrac{1}{2}D_n^2(\mathbf{z})$$

Note that $LLR(\mathbf{z}) = S_n(\mathbf{z}) - S_\gamma(\mathbf{z})$. In later sections, the two S-scores will be used to score anomalies.

Labeled data can be used to optimize performance ([11] describes a method of labeling PSD neutrons and gammas). With labeled data, the user compares the likelihood ratio to a threshold that is tuned to give a desired false positive rate (FPR). In the absence of labeled data, the performance can be tuned using probabilities estimated from the assumption that the fitted Gaussians correctly model the training data, otherwise, the FPR and related measures like true positive rate (TPR) are more important quantities than probabilities based on an idealized model, as in Equation (2).

*3.2 GMM and anomaly rejection*

All Users of PSD-based detector systems encounter problems trying to classify events that are neither neutrons nor gammas, called anomalies. Of special concern is that, in high flux environments, anomalies are dominated by pileups, which in turn are dominated by piled up gamma events. Finally, the pileups, especially in organic scintillators, preferentially contaminate the events classified as neutrons.

The main problem with pileups is that, unlike neutrons and gammas, they are not characterized by a consistent set of shapes. Instead of simply being one of two kinds of broader or narrower pulses, pileups are liable to have bumps and shoulders anywhere along the leading edge or tail of the pulse. This means that the anomaly problem is less about matching a definable type than to reject events that are different from known classes.

This work reports on the effectiveness of methods using a GMM trained only on neutrons and gammas to reject anomalous pulses from pulses classified as neutrons. Judging effectiveness requires tests with known pulse types. Assembling a collection of known pileup pulse types takes some effort because, unlike gamma and neutron pulses, real experimental pileups are very hard to label properly.

*3.2.1 Splitting neutrons from gammas*

Our foremost concern is with the anomalies that are misclassified as neutrons. We will limit the remaining discussion to those that have been classified by the original GMM as neutrons, and describe and test methods to separate the pileups from the neutrons. To restrict the investigation to events that would be classified as neutrons, we make a first cut using the optimal properties of LLR, as noted above. We set the threshold for neutrons at LLR = 0. This is the point where the likelihoods of a pulse being a neutron or a gamma are equal, as determined by the GMM. Among the remaining events with LLR > 0, we will compare the behaviour of cuts between the neutrons and pileups. The three-step method of finding a pulse, accepting neutrons, and rejecting anomalies is shown in Figure 1.

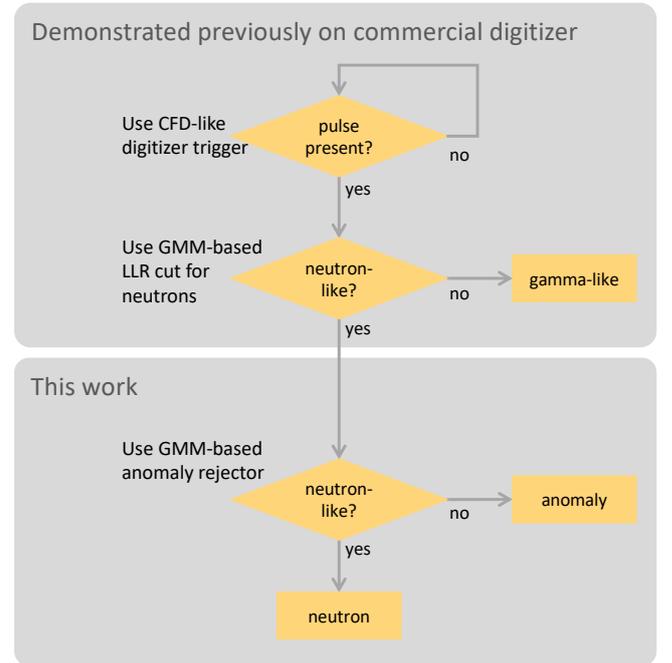

**Figure 1**. Flow chart for finding and classifying electronic pulses read out of gamma-neutron PSD materials.

*3.2.2 Recruiting gamma-neutron GMM scores to make cuts between neutrons and anomalies*

Even as the machine learning field is continually making advances, judgement and deliberation are required to select an anomaly rejection approach suited to a specific task (see, e.g. [16]).

In the situation at hand, the GMM provides two scores. In the space of the two scores, we can plot training data (described in Section 4) containing the three kinds of events – neutrons, gammas, and pile-ups—, visually look for boundaries that separate them, and look for ways to draw those boundaries in a simple way. The left panel of Figure 2 shows LLR versus energy for Cf-252, Cs-137, and synthetic pileup. It seems clear that LLR and energy do not provide a means to separate pileup events from neutrons and gammas. Because we constrained our practical application to work only with unlabeled data, we limit our options to prefer either cuts on statistically rigorous parameters or, failing that, straight cuts in a space that is easy to compute from the mixture model $\{\theta_n, \theta_\gamma\}$.

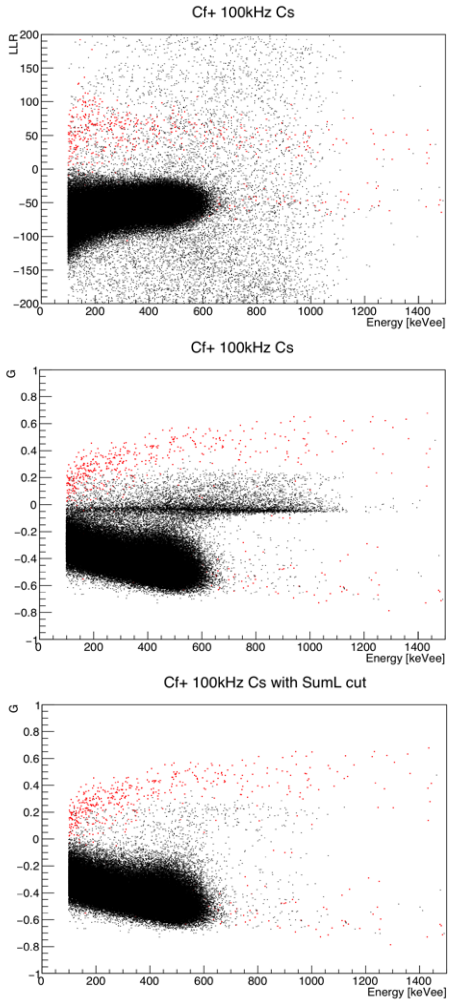

**Figure 2**. Cf-252 data (red points) combined with Cs-137 data (black points), including synthetic pileup in the amount expected to represent 100kHz of Cs-137 interactions. Top, the LLR score provides little ability to reject pileup. Center, the G score provides clearly improved pileup rejection, and gamma contamination of the neutron band drops drastically for G greater than ~0.25. Bottom, a cut on SumL drastically reduces pileup counts, but many with the most neutron-like scores remain. Full-color version available on-line.

### 3.2.3 Sum of likelihoods

The most straightforward and statistically rigorous test for an anomaly using the likelihoods provided by a two-component GMM is to sum the likelihoods. The derivation is as follows. That the observed event $z$ is an anomaly or not depends on the posterior probability of belonging to class neutron or class gamma given $z$, $P(class = n \cup \gamma | z)$. The smaller this probability, the more likely $z$ is an anomaly. It can be rewritten as being proportional to the sum of the two likelihoods under neutron and gamma, respectively.

$$P(class = n \cup \gamma | z) = P(class = n | z) + P(class = \gamma | z)$$
$$\propto P(class = n, z) + P(class = \gamma, z)$$

$$= P(z|class = n)P(class = n)$$
$$+ P(z|class = \gamma)P(class = \gamma)$$
$$= \alpha_n f_n(z) + \alpha_\gamma f_\gamma(z)$$

(In this notation, $P(A|B, C)$ means "the probability of A given B and C"). Note that in the special case when $\alpha_n = \alpha_\gamma$, the above posterior probability is proportional to the sum of the likelihoods. The likelihood of observing $z$ given a class, and the prior on the class probabilities, are estimated from the training data set. This technique is known as Bayesian network based semi-supervised anomaly detection [14].

In practice, the sum of likelihoods can be difficult to compute because the likelihoods are the constant $e$ to a large negative number. In order to make a practical computation of the log of the sum of the likelihoods, the exact expression, where $K_n = K_\gamma = K_{shared}$, is:

$$\ln(f_n(z) + f_\gamma(z))$$
$$= \ln(\exp(S_n(z)) + \exp(S_\gamma(z))) + K_{shared}$$

leading to

$$\ln(\exp(S_n) + \exp(S_\gamma))$$
$$= -\frac{1}{2}\min(S_n, S_\gamma) + \ln(1 + \exp(-|S_n - S_\gamma|))$$

In the region of neutron events provided by the LLR = 0 cut, the log sum of likelihoods is almost a straight cut using only the $S_n$ score.

### 3.2.4 Convenient expression of ratio of log-likelihoods

This work also proposes another parameter based on the likelihoods. We call this parameter G, and it is formed by dividing the negative of the sum of the *S*-scores out of the LLR for $z$ (the negative sum of S-scores makes G positive when LLR is positive, and negative when LLR is negative). As noted above, $K_n = K_\gamma$ in our choice of covariance structures, so G is:

$$G(z) = \frac{S_\gamma(z) - S_n(z)}{S_n(z) + S_\gamma(z)}$$

Cuts in this parameter are the same as those along the parameter that is the ratio of S-scores. Although G is a simple transformation of the ratio of the S-scores,

$$G = 1 - \frac{2}{1 + \frac{S_\gamma}{S_n}}$$

it has two particularly convenient properties: 1) G = 0 at LLR = 0 and 2) G is more well-behaved than the ratio of scores, running from -1 to 1 instead of 0 to infinity. The properties of G with shared covariance are such that a pulse at the mean of the neutron Gaussian component will have G of 1, at the mean of the gamma will have a G of -1, and anomalies will have G near zero. In a non-rigorous application, a plot of G versus energy allows the user to set by-eye cuts to suppress anomalies or as guided quantitatively by synthetic pileup discussed in Section 4.2. Figure 2 shows the qualitative comparison of separating events of different types using LLR, G, and G with a cut on sum of likelihoods. Sections 4.4 and 4.5 describe how to use synthetic pileups to guide the cuts quantitatively.

## 4. Tests

When labeled data is available, performance of several approaches can be compared using trade curves. In the present investigation, lab-acquired labeled pileups are hard to come by, instead we synthesize sets of some expected cases of pileups and use them to evaluate the performance of methods that use the scores from the GMM. Because this investigation occurs only in the region of S scores cut by LLR to favor neutrons, the neutron test set is also labeled in an informal manner. The neutron datasets for training and test will be all events from our low-rate Cf-252 data with GMM-derived LLR > 0.

Performance comparisons will be made using the trade curves known as ROC curves. ROC curves show the tradeoff between properly classifying neutrons as neutrons, and misclassifying pileups as neutrons. We will also compare performance using precision-recall (P-R) curves, a trade curve that takes the relative rates of neutrons and pileups into account to determine the fraction of events classified as neutrons that really are neutrons. P-R is important in situations where the ratio of all pileups to all neutrons is high even if the fraction of pileups misclassified as neutrons is low.

In the GMM score space, the neutrons have a large overlap with anomalies. In order to show the limitation imposed by this challenging situation, we will compare the performance of our feasible approaches against methods that are outside the constraints imposed by our application, namely we will obtain performance curves for a few conventional supervised methods. These supervised methods should outperform the unsupervised methods and benchmark the limitations of working with data with a large overlap. Computationally simple metrics that are easily compatible with FPGA implementation are of particular interest.

Tests with ROC curves and supervised training can only be performed when one has well-labeled data, where the instances are strictly of one class or the other. The synthetic pileups are strictly pileup data, although subject to any mismatch between synthesized and true pileup events. The LLR > 0 Cf-252 data is not strictly fast neutrons, it has a small amount of contamination from gamma events. For ROC curves, contamination in the "positives", which in this case are neutrons, means that the true positive rate will be artificially low especially near the transition between low-TPR/low-FPR and high-TPR/high-FPR. To determine the approximate size of the effect of gammas contaminating the Cf-252 neutrons used as ground truth, we made the following estimate. For pure Cs-137 gammas, we built the LLR distribution with binwidth of 1. The ratio of counts in the peak bin to number of Cs-137 events with LLR > 0 is about 2.6. For the gamma-and-neutron Cf-252 source, we also built the LLR distribution with binwidth of 1. The Cf-252 distribution has two peaks, and the peak below LLR = 0 contains mostly gammas. If the LLR < 0 gamma peak is in the same ratio to LLR > 0 gammas for both Cf-252 and Cs-137, then we can estimate the number of gammas contaminating the LLR > 0 Cf-252 events and compare that number with the total number of LLR > 0 Cf-252 events. The ratio is 1:200, meaning that we estimate no more than 1/200 events used for obtaining the TPR is actually a contaminating gamma event. Bear in mind that this small effect will reduce the apparent TPR for a given FPR for all the ROC curves approximately the same and so will have very little influence on the comparison among ROC curves.

### 4.1 Experimental set up and training the GMM

The experimental setup consisted of a 4-inch diameter 3-inch long EJ-309 cell read-out by a Hamamatsu R6233 PMT into a Struck 3316 digitizer card running at 250 Msamples/sec and collecting pulses in a 256-sample window with a 40-sample pre-trigger. To do the pulse processing, the 256 samples are split into an 8-sample pre-trigger baseline and a 128-sample pulse window. The trigger threshold corresponded to approximately a 10 keV electron.

We collected two datasets: Cf-252 at 692 events per second and Cs-137 at 30,442 events per second. The Cs-137 dataset was used for single-point energy calibration. The Cf-252 dataset consists almost entirely of non-anomaly neutron events and gamma events. This dataset was split into two unequal parts. The first 2000 events were used to train the two-component GMM. Only pulses of approximately 80 to 3000 keVee, the range near the synthetic trigger threshold discussed below to the highest energy non-saturated events, were used in the training. The rest of the Cf-252 data is used to provide neutron events for testing, with a LLR > 0 and energy > 100 keVee cut to remove nearly all the gammas. The Cs-137 dataset, consisting almost entirely of gamma events and piled up gamma events, is used to synthesize piled-up gamma events. Figure 3 shows the rate of contamination of non-neutron events from the LLR > 0 cut.

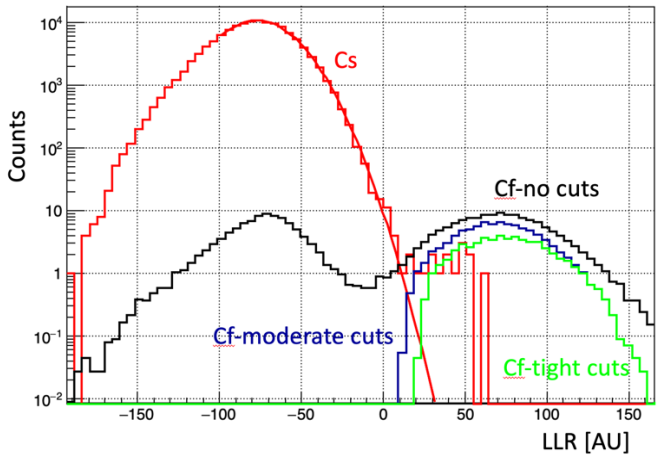

**Figure 3**. The LLR distributions for low rate Cs-137 (red) and Cf-252 (black) data. The Cf distribution is scaled down to approximate a 1:1000 neutron signal to gamma background. For the most basic purity cuts, pileup background is expected to dominate over misidentification of single pulses, represented by the smooth distribution. The excellent approximation of the Cs LLR distribution by a Gaussian for LLR > -100, as shown by the red curve, indicates that pileup is likely to be the dominant background for even a 1:10000 neutron signal to gamma background for tight cuts in G that cause ~50% efficiency loss, as shown by the green curve. Full-color version available on-line.

*4.2 Synthesizing pileup events*

In order to train and/or test with labeled pileups, we rely on synthesizing piled-up gamma pulses. A data set of 662 keV gamma interactions was collected from Cs-137. The rate of events in the detector was 30k/s. The rate of background fast neutrons in our detector is approximately less than one per second, so the contamination rate of fast neutrons in this gamma dataset of a few seconds can be estimated informally as fewer than 10 fast neutrons per 200,000 gamma pulses.

Pulses were synthesized as follows:

1. An 80 keVee threshold, significantly higher than the original near-noise trigger threshold, is selected.
2. An event is drawn from the low threshold Cs-137 dataset. If it is greater than the threshold, it is kept as the trigger pulse.
3. If it is kept, a second event is drawn.
4. The second event is baseline-subtracted.
5. If the second event is less than the threshold, it is time-shifted so that its peak lies anywhere in the window, positive or negative, and added to the trigger pulse.
6. If the second event is greater than the threshold, it is time-shifted a non-negative offset (including zero) and added to the trigger pulse.
7. If the summed event has a calculated energy greater than a 100 keVee threshold, it is saved as a synthetic pileup in a dataset of synthesized piled-up Cs-137 pulses.

Step 6 takes care of the case where a negative time-offset of a pulse above the threshold would not simulate a real pile-up case, because the leading pulse would cause a trigger for itself instead of for the trailing pulse. This set of synthetic data provides us with a set of pulses known to be piled-up gamma pulses with known offsets. Approximately 1 million pileup Cs-137 pulses were synthesized that way. Using the GMM trained on the Cf-252 dataset, the two S-scores are computed for each of these synthetic pileup pulses.

*4.3 Supervised benchmarks*

For this and previous papers, the authors' practical application required ruling out supervised training; however, we can implement supervised training to characterize the limits to performance imposed by the overlap of GMM scores between neutrons and pileups. Supervised learning requires a labeled dataset. The instances labeled as neutrons come from the Cf-252 data while the instances labeled as non-neutrons come from the synthetic pileup data. Training and test are applied only to the subset of these two datasets for which LLR > 0. The features consist of the neutron and gamma S-scores calculated using GMM on Cf-252 data. For some approaches we used the energy as a third feature. We used MATLAB Classification Learner App to train various classifiers and applied these classifiers to all the data to create ROC curves. We selected a range of supervised approaches including tree, several KNNs, ensemble, etc. The KNNs were built for three cases: K = 1, 10, and 100 nearest neighbors. Both two (S-scores) and three (S-scores plus energy) features were considered for various classifiers with 5-fold cross-validation.

*4.4 Performance testing using ROC curves*

We computed ROC curves for all approaches. We also selected the area under the curve (AUC) as a single summary statistic to compare overall performance of all the approaches. AUC essentially provides the probability that a classifier will rank a randomly chosen neutron (positive) instance higher than a randomly chosen pileup (negative) instance. Figure 4 shows the ROC curves for the best of the supervised methods and the anomaly-rejector scores described in Sections 3.2.3 and 3.2.4. Table 1 shows the AUC for all the approaches. KNN for K = 100 using three features gives the best performance. The overall unsupervised performance is the convex hull of the two ROC curves using thresholds on the sum of likelihood and the G parameter, respectively, as any point on the line segment between two prediction results can be achieved through the randomization technique [15]. One randomly uses one or the other system with probabilities proportional to the relative length of the opposite component of the segment. The unsupervised scores compare well to the supervised approaches using two features. Improvements in the KNN ROC curve by using energy as a third feature, denoted as "KNN3", suggests that much can be gained, except for cases of very stringent pileup rejection, not only if the training is supervised (using a cut on LLR > 0 and by building synthetic pileups) but also if energy is used as an additional feature.

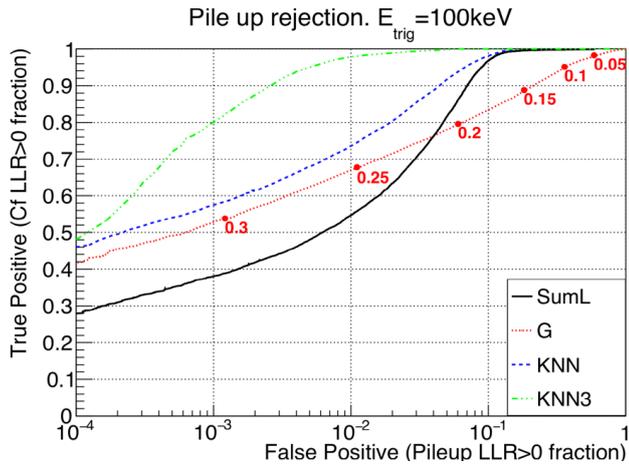

**Figure 4**. ROC curves for the sum of likelihoods (SumL black), G (red), KNN (blue) and KNN3 (green) scores. SumL and the G parameter are both derived from the two-component Gaussian mixture model of neutrons and gammas. The two K-nearest-neighbor methods require labeled data for supervised training, KNN3 uses energy as a feature in addition to the S-scores. "True positives" are neutrons with a positive LLR derived from a gamma-neutron GMM, and "false positives" are LLR > 0 synthetic pileups misidentified as neutrons. Neutrons are from Cf-252 generating > 100 keVee of light output. SumL is a good score for the low pileup scenarios, eliminating 90% of neutron-like pileup with only a few percent efficiency loss. The G score provides better performance when pileup needs to be reduced by more than two orders of magnitude. Scoring based on supervised machine learning methods, such as KNN, show the potential for even greater performance, and KNN3 shows the performance gains from including energy dependence. However, the performance gains are small in comparison to the analytical G score for the highest pileup rejection (lowest false positive rates) scenarios. The G score thresholds are notated at six points along the curve. Full-color version available on-line.

*4.5 Performance testing using Precision-Recall curves*

For many practical applications, the neutrons may be overwhelmed by non-neutron particles. In these cases, a small fraction of non-neutrons misidentified as neutrons (low FPR) may be the majority of events classified as neutrons, and a precision-recall curve (P-R) is more directly informative than an ROC (If the population ratio between classes is known, P-R can be computed from a ROC: recall is the TPR, and precision is a function of TPR, FPR, and the population ratio). A P-R curve calls attention to the question of how much the pulses classified as neutrons are diluted by non-neutrons. In this application, the precision-recall curve may be thought of as a trade between "neutron purity" and "neutron retention".

**Table 1**
Comparison of performance of ten supervised and two unsupervised classification methods. The figure of merit is the area under the ROC curve, AUC. The supervised classifiers are three K nearest-neighbor models with different values of K, a tree, and boosted trees. Each supervised classifier was trained with neutron and pileup labeled data, using as input features just the two parameters from the neutron-gamma GMM, and also the two GMM features plus the energy.

| | | AUC | |
|---|---|---|---|
| | | (E,Dn,Dg) | (Dn,Dg) |
| **SumL** | | n/a | .98 |
| **G** | | n/a | .94 |
| **KNN** | K = 1 | .98 | .91 |
| | K = 10 | .99 | .98 |
| | K = 100 | 1.0 (KNN3 in Figures) | .99 (KNN in Figures) |
| **Tree** | max splits = 100 | .99 | .98 |
| **Ensemble** | Boosted trees | 1.0 | .98 |

As with the ROC curves, we performed P-R on events in the LLR > 0 region. Precision depends on the ratio of neutrons versus everything else. Figure 3 shows that ignoring direct single gamma pulse misidentification is sufficient for our case, but handling cases where this approximation is not true may be accomplished by using a large sample of low rate data to estimate the single gamma pulse misidentification contribution. Figure 5 shows P-R for relative amounts of 25 Cs-137 pileups per Cf-252 neutron. As shown in the title of the plot, 25 pileups per neutron amounts to a Cs-137 rate of 10 kHz and neutron rate of 2.2 Hz, or a Cs-137 rate of 100 kHz and a neutron rate of 220 Hz; note the nonlinear scaling. Under these conditions, the unsupervised parameter G can be used to provide nearly 100% pure neutrons with an efficiency of 60% of all neutrons, while the sum of likelihoods can be used to provide nearly 100% neutron efficiency with 60% pure neutrons.

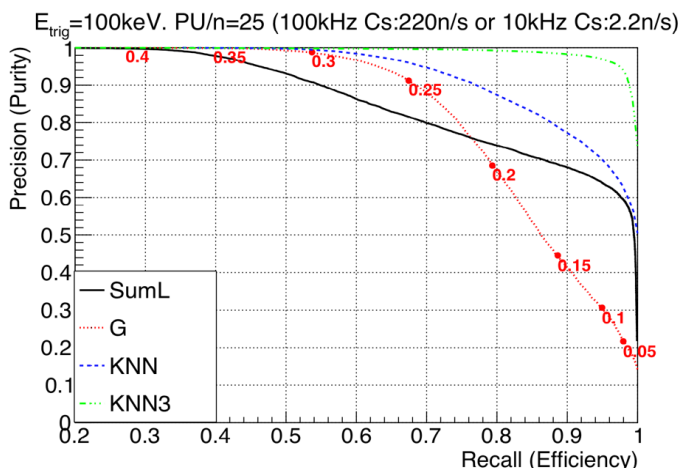

**Figure 5**. Precision-recall curves for neutrons from Cf-252 for the SumL (black), G (red), KNN (blue) and KNN3 (green) scores. Assumes misidentification is due only to Cs-137 pileup for scenarios with 25 pulse pileups per detected neutron (i.e. creating > 100 keVee of light output). SumL is a good score if efficiency above 80% is required. The G score provides better performance when high purity is needed at the cost of at least 30% signal loss. Scoring based on supervised machine learning methods, such as KNN, show the potential for even greater performance, and KNN3 shows the performance gains from including energy dependence. However, the performance gains are small in comparison to the analytical G score for the highest purity scenarios. The G score thresholds are notated at eight points along the curve. Full-color version available on-line.

## 5. Discussions and Conclusions

This work describes a way of taking advantage of a two-component GMM to build an anomaly rejecter. In this study, anomaly rejecters are designed and applied specifically to gamma-neutron pulse shape discriminating scintillators. For this problem, piled-up pulses contaminate the LLR-discriminated neutron component of the GMM, while at the same time they do not make up a well-constrained cluster in feature space. Statistical rigor provides one solution, the sum of likelihoods, and the specific PSD situation constrains the distributions of anomalies well enough to provide another practical unsupervised solution via the G score. The G score is particularly valuable for situations requiring several orders of magnitude of pileup rejection.

Operationally, the method leaves the user with three kinds of events: neutrons, LLR > 0 anomalies consisting of pileups and low-signal-to-noise neutrons, and LLR < 0 non-neutrons that are almost entirely gammas and piled-up gammas. The losses of neutrons in the overlap region and at LLR < 0 are folded into the efficiency of the system. In cases where either high gamma background or pileup is a problem, this method of rejecting anomalies both improves the efficiency by separating neutrons from other particles, and reduces the problems associated with including non-neutrons as neutrons in the analysis of data. For example, the analytical G score improves the presented scenario, over using only the LLR cut, from 16% neutron purity to nearly 99.9% while incurring a 60% loss over the baseline neutron detection efficiency. Comparatively, a direct reduction of detector size to 60% lower efficiency would increase the purity to approximately only 32%. We expect many applications dependent on fast neutron measurements, such as active interrogation and neutron multiplicity counting will benefit from the application of these techniques.

*Declaration of Competing Interest*

The authors declare that they have no known competing financial interests or personal relationships that could have appeared to influence the work reported in this paper.

*Acknowledgements*

We would like to thank Les Nakae and Sean Walston for their support in the preparation of this paper. We would like to thank Brenton Blair for helpful discussions. This work was performed under the auspices of the U.S. Department of Energy by Lawrence Livermore National Laboratory under Contract DE-AC52-07NA27344. We would like to thank the U.S. Department of Energy, National Nuclear Security Administration, Office of Defense Nuclear Nonproliferation Research and Development and the Office of Nonproliferation and Arms Control. This support does not constitute an express or implied endorsement on the part of the government.